\documentclass[oneside,12pt,a4paper]{article}

\usepackage[ansinew]{inputenc}
\usepackage{booktabs, tabularx}
\usepackage{epsfig}
\usepackage{float}
\usepackage{picins}
\usepackage{amsmath}
\usepackage{amsfonts}
\usepackage{amssymb}
\usepackage{enumerate}
\usepackage{afterpage}
\usepackage{textcomp}
\usepackage{setspace}
\floatstyle{plain}

\begin{document}
\begin{center}
  {\Large \bf ASN-Minimax double sampling plans by variables for two-sided specification limits when $\sigma$ is unknown}\\[0.5cm]
   {\sc Eno Vangjeli}\\[0.3cm]
\end{center}
\vspace*{0.1cm}  {\small }
{\small{\bf{Abstract:}} ASN-Minimax double sampling plans by variables for a normally distributed quality characteristic with unknown $\sigma$ and two-sided specification limits are introduced. These plans base on the essentially Maximum-Likelihood (ML) estimator $p^*$ and the Minimum Variance Unbiased (MVU) estimator $\hat{p}$ of the fraction defective $p$. The operation characteristic (OC) of the ASN-Minimax double sampling plans is determined by using the independent random variables $p^*_1$, $p^*_2$ and $\hat{p}_1$, $\hat{p}_2$, which relate to the first and second samples, respectively. The maximum of the average sample number (ASN) of these plans is shown to be considerably smaller than the sample size of the corresponding single sampling plans.\\\\
{\em Keywords:} Acceptance sampling by variables, ASN-Minimax double sampling plan, essentially Maximum-Likelihood estimator, Minimum Variance Unbiased estimator}\\
\begin{center}
1. {\sc Introduction}
\end{center}
\vspace*{2mm}
For a normally distributed characteristic $X \sim N(\mu, \sigma^2)$ with unknown $\sigma>0$ and lower and upper specification limits $L$ and $U$, respectively, the fraction defective function $p(\mu, \sigma)$ is defined as
\begin{equation}
p(\mu, \sigma):=P(X<L)+P(X>U)=\Phi\left({L-\mu\over \sigma}\right)+\Phi\left({\mu-U\over \sigma}\right),
\end{equation}
where $\Phi$ denotes the standard normal distribution function. Given a large-sized lot and a single sample $X_1,...,X_n$, $(n>3)$ with
\begin{equation*}
\overline{X}={1\over n}\sum_{i=1}^{n}X_i, \quad  S^2={1\over n-1}\sum_{i=1}^{n}(X_i-\overline{X})^2,
\end{equation*}
there are two well-known procedures to determine single sampling plans. The first procedure (cf. {\sc Bruhn-Suhr}/{\sc Krumbholz} (1990)) is based on the essentially Maximum-Likelihood (ML) estimator
\begin{equation}
p^*=p(\overline{X}, S)=\Phi\left({L-\overline{X}\over S}\right)+\Phi\left({\overline{X}-U\over S}\right).
\end{equation}
The second procedure (cf. {\sc Bruhn-Suhr}/{\sc Krumbholz} (1991)) is based on the Minimum Variance Unbiased (MVU) estimator $\hat{p}$ (cf. {\sc Kolmogorov} (1953), {\sc Lieberman}/{\sc Resnikoff} (1955)). Let
\begin{equation}
V:=max\biggl\{0, {1\over 2}-{1\over 2}{\overline{X}-L\over S}{\sqrt{n}\over n-1} \biggr\},
\end{equation}
\begin{equation}
W:=max\biggl\{0, {1\over 2}-{1\over 2}{U-\overline{X}\over S}{\sqrt{n}\over n-1} \biggr\},
\end{equation}
\begin{equation}
b(x):={\Gamma(n-2)\over\displaystyle\Gamma\left({n-2\over 2}\right)\Gamma\left({n-2\over 2}\right)}\>x^{{n \over 2}-2}\>(1-x)^{{n \over 2}-2} \quad (0<x<1),
\end{equation}
\begin{equation}
B(x):=\int_0^xb(t)dt,
\end{equation}
where $\Gamma$ denotes the gamma function and $B(x)$ and $b(x)$ the distribution and density function of the symmetrical beta distribution with parameter $\displaystyle {n-2\over 2}$. The MVU estimator $\hat{p}$ is defined by
\begin{equation}
\hat{p}:=B(V)+B(W).
\end{equation}
The lot is accepted within the single sampling plan
\begin{eqnarray*}
&&(n^*, ~k^*), \quad \text{if} \quad p^*\leq k^* \quad \text{for the first procedure},\\
&&(\hat{n}, ~\hat{k}), \hspace*{7.5mm} \text{if} \quad \hat{p}\leq \hat{k} \hspace*{7.5mm} \text{for the second procedure}.
\end{eqnarray*}
In this paper, we introduce ASN-Minimax (AM) double sampling plans for both procedures. For given acceptable quality level $p_1$, rejectable quality level $p_2$ and levels $\alpha$ and $\beta$ of Type-I and Type-II error, respectively, the AM-double sampling plans feature minimal maximal average sample number (ASN) while satisfying the classical two-points-condition on the operation characteristic (OC). We define the double sampling plans by using in the second stage only the information obtained from the second sample. Thus, the double-sampling-plan-OC can be determined by implementing the corresponding single-sampling-plan-OC. The double sampling plans are computed in a similar way as the single sampling plans.\\
In the next section, we give two well-known theorems regarding the single-sampling-plan-OC for each procedure and derive with their help the corresponding double-sampling-plan-OC. The computation of the AM-double sampling plans is elucidated in the third section. In order to determine two sample sizes and three critical values of the two-sided AM-double sampling plan, we use the corresponding one-sided AM-approximation.\\
\begin{center}
2. {\sc The double-sampling-plan-OC}
\end{center}
\vspace*{3mm}
Before deriving the double-sampling-plan-OC, we should bring together two well-known theorems from the literature on single sampling. Let
\begin{subequations}
\begin{equation}
 L_{(n^*, ~k^*)}(\mu, \sigma)=P(p^*\leq k^*)
\end{equation}
and
\begin{equation}
 L_{(\hat{n}, ~\hat{k})}(\mu, \sigma)=P(\hat{p}\leq \hat{k})
\end{equation}
\end{subequations}
denote the single-sampling-plan-OC for the first and second procedures, respectively, and let $g_r$ be the density function of the $\chi^2$ distribution with $r$ degrees of freedom.\\\\
{\bf Theorem 1:} \quad It holds that (cf. {\sc Bruhn-Suhr}/{\sc Krumbholz} (1990)):
\begin{eqnarray}
&& L_{(n^*, ~k^*)}(\mu, \sigma)=\int_0^C\biggl\{  \Phi\left({\sqrt{n^*}\over\sigma}\left(\mu\left(\sigma \sqrt{{t \over n^*-1}}, k^* \right)-\mu\right)\right)\nonumber\\
&&-\Phi\left({\sqrt{n^*}\over\sigma}\left(\acute{\mu}\left(\sigma \sqrt{{t \over n^*-1}}, k^* \right)-\mu\right)\right) \biggr\}g_{n^*-1}(t)dt
\end{eqnarray}
with
\begin{equation*}
C={(n^*-1)(L-U)^2 \over 4\sigma^2\left(\Phi^{-1}\displaystyle\left({k^*\over2}\right)\right)^2} \quad \text{and} \quad \acute{\mu}(\sigma, p)=L+U-\mu(\sigma, p).
\end{equation*}\\
{\bf Theorem 2:} \quad It holds that (cf. {\sc Bruhn-Suhr}/{\sc Krumbholz} (1991)):
\begin{eqnarray}
&& L_{(\hat{n}, ~\hat{k})}(\mu, \sigma)=\int_0^{A(B^{-1}(\hat{k}), 0)}\biggl\{  \Phi\left(-\delta_U - \sqrt{t(\hat{n}-1)} \right)-\nonumber\\
&&-\Phi\left( -\delta_L + (1-2B^{-1}(\hat{k})) \sqrt{t(\hat{n}-1)}   \right) \biggr\}g_{\hat{n}-1}(t)dt+\nonumber\\
&&+2\int_0^{B^{-1}(\hat{k})}\biggl\{ \int_0^{A(\psi(y), y)}    \Phi'\left(-\delta_U +(2y-1) \sqrt{t(\hat{n}-1)} \right)\cdot\nonumber\\
&&\cdot\sqrt{t(\hat{n}-1)}g_{\hat{n}-1}(t)dt\biggr\}dy
\end{eqnarray}
with\\
\begin{equation*}
A(x,y)=
\begin{cases}
\displaystyle{\hat{n}(U-L)^2 \over 4\sigma^2(\hat{n}-1)\left( 1-x-y\right)^2}& \text{if} \quad x+y<1\\
\hspace*{2cm}\infty & \text{else},
\end{cases}
\end{equation*}
\vspace*{4mm}
\begin{equation*}
\delta_U=\sqrt{\hat{n}}{\mu-U\over \sigma}, \hspace*{.5cm} \delta_L=\sqrt{\hat{n}}{\mu-L\over \sigma} \quad \text{and} \quad \psi(y)=B^{-1}(\hat{k}-B(y)).
\end{equation*}\\
{\bf Definition 1:} \quad The double sampling plan by variables
\begin{eqnarray*}
&&\lambda^* = \left(\begin {array}{c c c}
n_1^* & k_1^* & k_2^*\\ n_2^* & k_3^*
\end {array}\right) \quad \text{for the first procedure},\\
&&\widehat{\lambda} = \left(\begin {array}{c c c}
\hat{n}_1 & \hat{k}_1 & \hat{k}_2\\ \hat{n}_2 & \hat{k}_3
\end {array}\right) \quad \text{for the second procedure},
\end{eqnarray*}
with $n_1^*,n_2^*,\hat{n}_1,\hat{n}_2\>{\in\mathbb{N}};\>\>n_1^*,n_2^*,\hat{n}_1,\hat{n}_2\geq 2;\>\>k_1^*,k_2^*,k_3^*,\hat{k}_1, \hat{k}_2,\hat{k}_3\>{\in\mathbb{R^{+}}};\>\>k_1^*\leq
k_2^*$,\\$\hat{k}_1 \leq \hat{k}_2$, is defined as follows:
\begin{enumerate}[(i)]
\item
Observe a first sample of size  $n_1^*$ for the first procedure, $\hat{n}_1$ for the second procedure and compute $p_1^*, \hat{p}_1$.\\\\
If $p_1^* \leq k_1^*$, $\hat{p}_1\leq \hat{k}_1$, accept the lot.\\
If $p_1^* > k_2^*$, $\hat{p}_1 > \hat{k}_2$, reject the lot.\\
If $k_1^* < p_1^* \leq k_2^*$, $\hat{k}_1<\hat{p}_1\leq \hat{k}_2$, go to (ii).
\item
Observe a second sample of size $n_2^*$ for the first procedure, $\hat{n}_2$ for the second procedure and compute $p_2^*, \hat{p}_2$.\\\\
If $p_2^* \leq k_3^*$, $\hat{p}_2\leq \hat{k}_3$, accept the lot.\\
If $p_2^* > k_3^*$, $\hat{p}_2 > \hat{k}_3$, reject the lot.\\\\
\end{enumerate}
\vspace*{-.7cm}
Let $p^{est} \in \{p^*, \hat{p}\} $ denote the used $p$-estimator, $(n,~k)\in \{(n^*,~k^*), (\hat{n},~\hat{k})\}$ the corresponding single sampling plan and $\lambda=\left(\begin {array}{c c c}
n_1 & k_1 & k_2\\ n_2 & k_3
\end {array}\right)$, $\lambda \in\{\lambda^*,\widehat{\lambda}\}$ the corresponding double sampling plan given in Definition 1. Considering the independence of the samples, the OC $L_{\lambda}(\mu, \sigma)$ is given by
\begin{eqnarray}
&&\hspace*{-15.9mm}L_{\lambda}(\mu, \sigma)=P_{(\mu, \sigma)}(p_1^{est} \leq k_1)+P_{(\mu, \sigma)}(p_2^{est} \leq k_3,~k_1 < p_1^{est} \leq k_2)\nonumber \\
&&=L_{(n_1, k_1)}(\mu, \sigma)+L_{(n_2, k_3)}(\mu, \sigma)\left(L_{(n_1, k_2)}(\mu, \sigma)-L_{(n_1, k_1)}(\mu, \sigma)\right).
\end{eqnarray}
The ASN $N_{\lambda}(\mu, \sigma)$ is given by
\begin{equation}
N_{\lambda}(\mu, \sigma)=n_1+n_2 \>P_{(\mu, \sigma)}(k_1 < p_1^{est} \leq k_2)
\end{equation}
with
\begin{equation*}
P_{(\mu, \sigma)}(k_1 < p_1^{est} \leq k_2)=L_{(n_1, k_2)}(\mu, \sigma)-L_{(n_1, k_1)}(\mu, \sigma).
\end{equation*}
Depending on $\mu$ and $\sigma$, $L_{\lambda}$ and $N_{\lambda}$ are bands in $p$ (Figures 1, 2). The maximum $N_{max}(\lambda)$ of $N_{\lambda}(\mu, \sigma)$ is for the computation of the AM-double sampling plans $\lambda_{AM}$ irrelevant. However, we calculate $N_{max}(\lambda)$ as a basis for comparison with the ASN maximum of the AM-one-sided approximation, which is used in the computation of $\lambda_{AM}$. A detailed description of the computation of $\lambda_{AM}$ is given in the next section.
\vspace{-10mm}
\begin{figure}[H]
\begin{center}
   \includegraphics[width=.99\columnwidth]{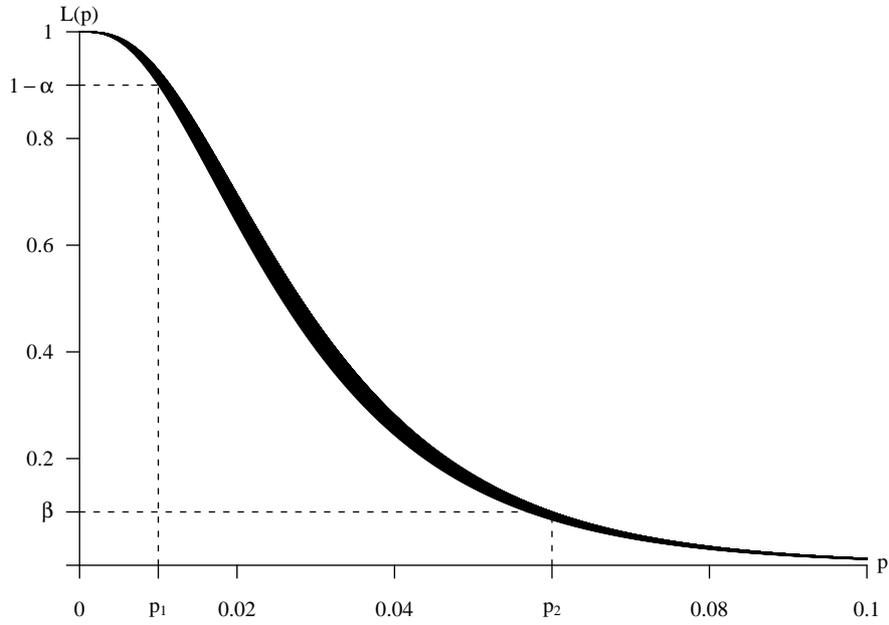}
   \vspace{-10mm}
    \caption{OC-band for $\lambda_{AM}^*$ defined by $p_1=0.01$, $p_2=0.06$ and $\alpha=\beta=0.1$}
    \label{ASN3}
\end{center}
\end{figure}
\vspace{-20mm}
\begin{figure}[H]
\begin{center}
   \includegraphics[width=.99\columnwidth]{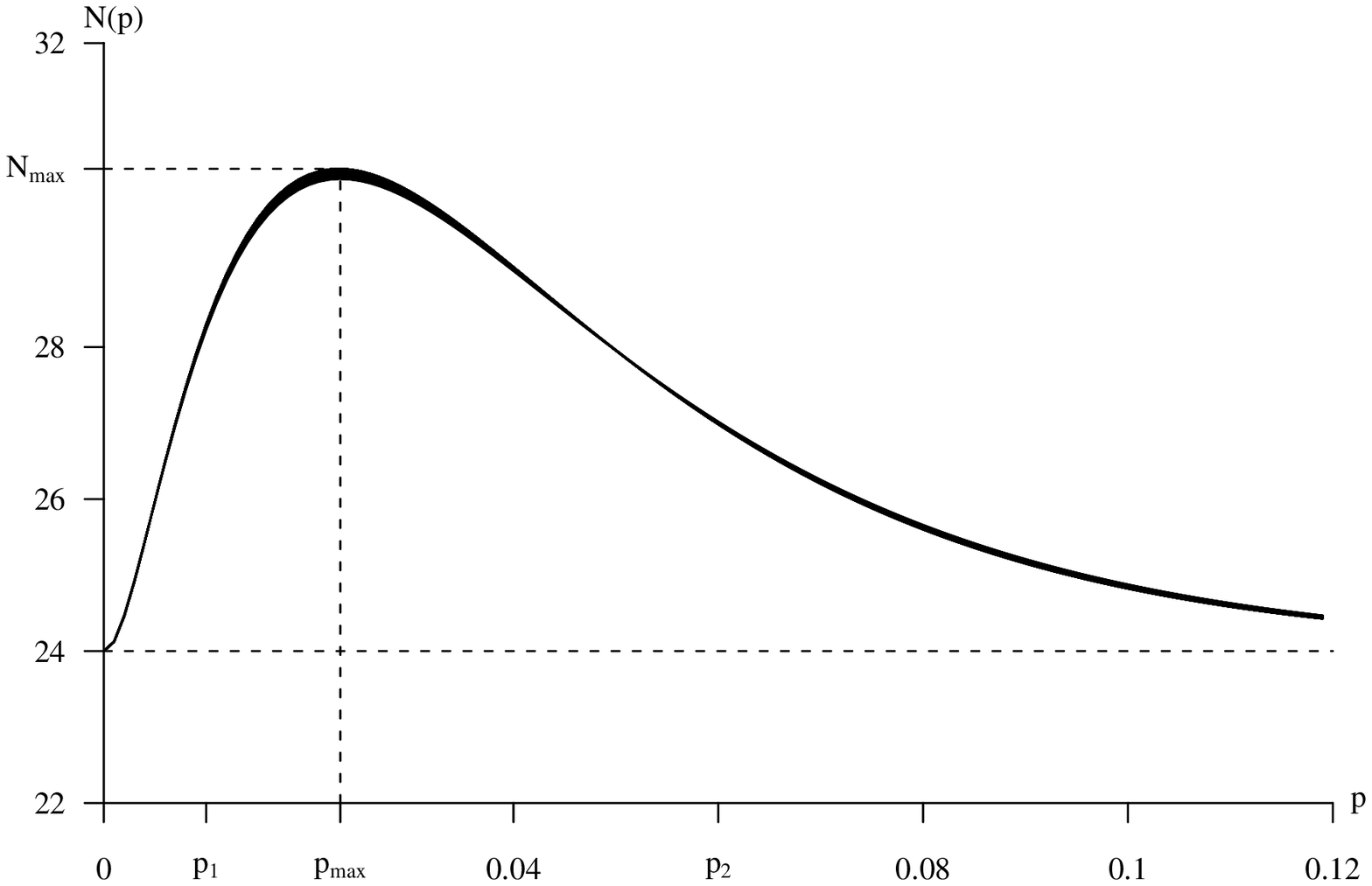}
   \vspace{-10mm}
    \caption{ASN-band for $\widehat{\lambda}_{AM}$ defined by $p_1=0.01$, $p_2=0.06$ and $\alpha=\beta=0.1$}
    \label{ASN3}
\end{center}
\end{figure}
\begin{center}
3. {\sc The computation of the AM-double sampling plans}
\end{center}
\vspace*{2mm}
The two-sided single sampling plan is calculated by using the one-sided approximation $(\tilde{n}, ~\tilde{k})$, which, without loss of generality is chosen for the case of an upper specification limit $U$. For given acceptable quality level $p_1$, rejectable quality level $p_2$ and levels $\alpha$ and $\beta$ of Type-I and Type-II error, respectively, the one-sided approximation $(\tilde{n}, ~\tilde{k})$ is computed by fulfilling the condition
\begin{eqnarray}
&\text{(i)}& F_{\tilde{n}-1,~\delta(p_1)}(l)\geq 1-\alpha\nonumber\\
&\text{(ii)}& F_{\tilde{n}-1,~\delta(p_2)}(l) \leq \beta\nonumber\\
&\text{(iii)}& \tilde{n}\stackrel{!}{=}\text{min}\\
&\text{(iv)}& \tilde{k}=
\begin{cases}
\Phi\left( \displaystyle{ l \over \sqrt{\tilde{n}}}\right)&\text{for the first procedure}\\
B\left( \displaystyle{ 1 \over 2}+{l \over 2(\tilde{n}-1)}\right) &\text{for the second procedure}.\nonumber
\end{cases}
\end{eqnarray}
$F_{\tilde{n}-1,~\delta(p)}$ denotes the distribution function of the noncentral t-distribution with $\tilde{n}-1$ degrees of freedom and the non-centrality parameter
\begin{equation}
\delta(p)=\sqrt{\tilde{n}}\>\Phi^{-1}(p).
\end{equation}
Let for the case of an upper specification limit $U$, $\phi^*=\left(\begin {array}{c c c}
n_1 & l_1 & l_2\\ n_2 & l_3
\end {array}\right)$ denote the AM-double sampling plan defined by using the independent statistics
\begin{equation}
T_i=\sqrt{n_i}\>{\overline{X}_i-U\over S_i}, \quad i=1, 2.
\end{equation}
The OC and ASN for a double sampling plan $\phi$ in this case (cf. {\sc Hilbert} (2005)) are given by
\begin{equation}
L_\phi(p)=F_{n_1-1,~\delta_1(p)}(l_1)+F_{n_2-1,~\delta_2(p)}(l_3)\>(F_{n_1-1,~\delta_1(p)}(l_2)-F_{n_1-1,~\delta_1(p)}(l_1))
\end{equation}
and\\
\begin{equation}
N_\phi(p)=N(p)=n_1+n_2\>(F_{n_1-1,~\delta_1(p)}(l_2)-F_{n_1-1,~\delta_1(p)}(l_1)),
\end{equation}
respectively, where $\delta_i(p)=\sqrt{n_i}\>\Phi^{-1}(p)$ for $i=1,2$.\\
$\phi^*$ is computed by
\begin{eqnarray}
&\text{(i)}&L_\phi(p_1)\geq 1-\alpha\nonumber\\
&\text{(ii)}& L_\phi(p_2) \leq \beta\\
&\text{(iii)}&N_{max}(\phi^*)= \min_{\phi \in Z} N_{max}(\phi),\nonumber
\end{eqnarray}
where $Z$ denotes the set of the double sampling plans $\phi$ fulfilling (18)(i) and (ii). The algorithm for the computation of $\phi^*$ is similar to that developed by {\sc Vangjeli} (2009) for computing AM-double stage tests. The parameters of the approximated AM-double sampling plan $\widetilde{\lambda}=\left(\begin {array}{c c c}
\tilde{n}_1 & \tilde{k}_1 & \tilde{k}_2\\ \tilde{n}_2 & \tilde{k}_3
\end {array}\right)$ in the two-sided case are given by $\tilde{n}_1=n_1$, $\tilde{n}_2=n_2$ for both procedures and
\begin{equation*}
\tilde{k}_1=\Phi\left( {l_1 \over \sqrt{n_1}}\right), ~ \tilde{k}_2=\Phi\left( {l_2 \over \sqrt{n_1}}\right), ~  \tilde{k}_3=\Phi\left( {l_3 \over \sqrt{n_2}}\right),
\end{equation*}
for the first procedure, and
\begin{equation*}
\tilde{k}_1=B\left( {1 \over 2}+{l_1 \over 2(n_1-1)}\right), ~  \tilde{k}_2=B\left( {1 \over 2}+{l_2 \over 2(n_1-1)}\right), ~ \tilde{k}_3=B\left( {1 \over 2}+{l_3 \over 2(n_2-1)}\right),
\end{equation*}
for the second procedure. Identically to the computation of the single sampling plans, we set
\begin{equation*}
\mu_0={L+U \over 2},
\end{equation*}
\begin{equation*}
\sigma_0(p)={L-U \over 2\Phi^{-1}\left( {p \over 2}\right)}
\end{equation*}
and define for given $p$ $(0<p<1)$ and $\sigma$ $(0<\sigma\leq\sigma_0(p))$, $\mu=\mu(\sigma, p)$ by inverting (1) with $\mu\geq\mu_0$. Setting
\begin{equation}
L_{\lambda_{AM}}(\sigma; p):=L_{\lambda_{AM}}(\mu(\sigma, p), \sigma),
\end{equation}
\begin{equation}
N_{\lambda_{AM}}(\sigma; p):=N_{\lambda_{AM}}(\mu(\sigma, p), \sigma),
\end{equation}
the AM-double sampling plan $\lambda_{AM}$ is given by the condition
\begin{eqnarray}
&\text{(i)}&\min_{0<\sigma\leq\sigma_0(p)}L_{\lambda_{AM}}(\sigma;p_1)\geq 1-\alpha\nonumber\\
&\text{(ii)}& \max_{0<\sigma\leq\sigma_0(p)}L_{\lambda_{AM}}(\sigma;p_2) \leq \beta \quad \text{(Figure 1)}\\
&\text{(iii)}&N_{max}(\phi^*)= \min_{\phi \in Z} N_{max}(\phi),\nonumber
\end{eqnarray}
the ASN maximum of $\lambda_{AM}$ is defined
\begin{equation}
N_{max}(\lambda_{AM})=\max\limits_{p,~\sigma}N_{\lambda_{AM}}(\sigma;p) \quad \text{(Figure 2)}.
\end{equation}
$\lambda_{AM}$ is calculated as follows:
\begin{enumerate}[1.]
\item
Compute the single sampling plan $(n,~k)$ and save $\alpha^*, \beta^*$
\item
Starting from $\alpha^{**}=\alpha^*, \beta^{**}=\beta^*$, compute $\widetilde{\lambda}$ for $p_1, p_2, \alpha^{**}, \beta^{**}$, verify (21)(i), (ii) and vary $\alpha^{**}, \beta^{**}$ until (21)(i), (ii) are true.\\
\end{enumerate}
\underline{Example 1}\\\\
For $~L=1$, $~U=9$, $~p_1=0.01$, $~p_2=0.06$, $~\alpha=\beta=0.1$, we get\\
\begin{center}
(i)\hspace*{6mm}$(n^*,~k^*)=(36, ~0.02645943143)~~$ and $~~\alpha^*=0.082, ~\beta^*=0.1$.
\end{center}
\vspace*{.5mm}
\begin{center}
\begin{tabular}{>{\footnotesize}c|>{\footnotesize}c|>{\footnotesize}c|>{\footnotesize}c|>{\footnotesize}c|>{\footnotesize}c}
$\alpha^{**}$ & $\beta^{**}$& $\widetilde{\lambda}$ &$N_{max}(\phi^*)$ &$\min\limits_{\sigma} L_{\widetilde{\lambda}}(\sigma;p_1) $& $ \max\limits_{\sigma}L_{\widetilde{\lambda}}(\sigma;p_2) $\\ \hline
0.082& 0.1& $\begin {array}{c c c} 25 & 0.016988 & 0.034411\\
19 & 0.029059\end {array}$ & 31.31538 & 0.8882933643 & 0.0999999893 \\
0.076& 0.1& $\begin {array}{c c c} 26 & 0.017688 & 0.034554\\
19 & 0.029215\end {array}$ & 32.16417 & 0.8960633610 & 0.0999999882\\
0.072& 0.1& $\begin {array}{c c c} 26 & 0.017577 & 0.035291\\
20 & 0.029275\end {array}$ & 32.75441 & 0.9010124424 & 0.0999999889\\
\end{tabular}
\end{center}
\vspace*{2mm}
where $\lambda^*_{AM}=\left(\begin {array}{c c c}
26 & 0.017577 & 0.035291\\ 20 & 0.029275
\end {array}\right)$ with $N_{max}(\lambda^*_{AM})= 32.75439$.\\
\begin{center}
(ii) \hspace*{5mm} $(\hat{n},~\hat{k})=(34, ~0.02262119182)~~$ and $~~\alpha^*=0.098, ~\beta^*=0.094$.
\end{center}
\begin{center}
\begin{tabular}{>{\footnotesize}c|>{\footnotesize}c|>{\footnotesize}c|>{\footnotesize}c|>{\footnotesize}c|>{\footnotesize}c}
$\alpha^{**}$ & $\beta^{**}$& $\widetilde{\lambda}$ &$N_{max}(\phi^*)$ &$\min\limits_{\sigma} L_{\widetilde{\lambda}}(\sigma;p_1) $& $ \max\limits_{\sigma}L_{\widetilde{\lambda}}(\sigma;p_2) $\\ \hline
0.098& 0.094& $\begin {array}{c c c} 24 & 0.011840 & 0.029264\\
18 & 0.023800\end {array}$ & 30.19915 & 0.9002805676 & 0.1004854552 \\
0.098& 0.093& $\begin {array}{c c c} 24 & 0.012148 & 0.029093\\
19 & 0.023424\end {array}$ & 30.34629 & 0.9002364848 & 0.0995258042 \\
\end{tabular}
\end{center}
where $\widehat{\lambda}_{AM}=\left(\begin {array}{c c c}
24 & 0.012148 & 0.029093\\ 19 & 0.023424
\end {array}\right)$ with $N_{max}(\widehat{\lambda}_{AM})= 30.34628$.\\\\
\underline{Example 2}\\\\
For $~L=1$, $~U=9$, $~p_1=0.01$, $~p_2=0.03$, $~\alpha=\beta=0.1$, we get
\begin{center}
(i)\hspace*{6mm}$(n^*,~k^*)=(115, ~0.0178762881)~~$ and $~~\alpha^*=0.085, ~\beta^*=0.1$.
\end{center}
\begin{center}
\begin{tabular}{>{\footnotesize}c|>{\footnotesize}c|>{\footnotesize}c|>{\footnotesize}c|>{\footnotesize}c|>{\footnotesize}c}
$\alpha^{**}$ & $\beta^{**}$& $\widetilde{\lambda}$ &$N_{max}(\phi^*)$ &$\min\limits_{\sigma} L_{\widetilde{\lambda}}(\sigma;p_1) $& $ \max\limits_{\sigma}L_{\widetilde{\lambda}}(\sigma;p_2) $\\ \hline
0.085& 0.1& $\begin {array}{c c c} 79 & 0.013777 & 0.021642\\
64 & 0.018624\end {array}$ & 101.1604 & 0.8948821204  & 0.0999999568 \\
0.083& 0.1& $\begin {array}{c c c} 80 & 0.013902 & 0.021726\\
64 & 0.018464\end {array}$ & 102.0913 & 0.8972199027  & 0.0999999565 \\
0.080& 0.1& $\begin {array}{c c c} 81 & 0.014029 & 0.021742\\
66 & 0.018537\end {array}$ & 103.5434 & 0.9008045948  & 0.0999999565 \\
\end{tabular}
\end{center}
where $\lambda^*_{AM}=\left(\begin {array}{c c c}
81 & 0.014029 & 0.021742\\ 66 & 0.018537
\end {array}\right)$ with $N_{max}(\lambda^*_{AM})= 103.5434$.\\
\begin{center}
(ii) \hspace*{5mm} $(\hat{n},~\hat{k})=(113, ~0.01678745123)~~$ and $~~\alpha^*=0.096, ~\beta^*=0.094$.
\end{center}
\begin{center}
\begin{tabular}{>{\footnotesize}c|>{\footnotesize}c|>{\footnotesize}c|>{\footnotesize}c|>{\footnotesize}c|>{\footnotesize}c}
$\alpha^{**}$ & $\beta^{**}$& $\widetilde{\lambda}$ &$N_{max}(\phi^*)$ &$\min\limits_{\sigma} L_{\widetilde{\lambda}}(\sigma;p_1) $& $ \max\limits_{\sigma}L_{\widetilde{\lambda}}(\sigma;p_2) $\\ \hline
0.096& 0.094& $\begin {array}{c c c} 78 & 0.012471 & 0.020036\\
62 & 0.017078\end {array}$ & 99.13899 & 0.9000170882 & 0.1014154024 \\
0.096& 0.092& $\begin {array}{c c c} 78 & 0.012406 & 0.020069\\
64 & 0.016981\end {array}$ & 100.1070 & 0.9000091667 & 0.0993767725 \\
\end{tabular}
\end{center}
where $\widehat{\lambda}_{AM}=\left(\begin {array}{c c c}
78 & 0.012406 & 0.020069\\ 64 & 0.016981
\end {array}\right)$ with $N_{max}(\widehat{\lambda}_{AM})= 100.1070$.
\newpage
\hspace*{-6.5mm}{\bf Remark:}
Numerical investigations have shown:
\begin{enumerate}[(i)]
\item
There are nonessential differences between $N_{max}(\phi^*)$ and $N_{max}(\lambda_{AM})$.
\item
Since the parameters $\alpha^{**}$ and $\beta^{**}$ vary by a constant of 0.001, we cannot presume that the calculated $\lambda_{AM}$ is the precise AM-double sampling plan. Finding plans that fulfill more sharply (21) (i), (ii) and feature an insignificantly lower $N_{max}$ would require an unreasonable amount of time.
\item
For the calculated double sampling plans, it holds that
\begin{equation*}
N_{max}(\widehat{\lambda}_{AM})<N_{max}(\lambda^*_{AM})
\end{equation*}
(Figure 3),
which is consistent with the known superiority of $\hat{p}$ over $p^*$ from the single sampling literature.
\end{enumerate}
\vspace*{-10mm}
\begin{figure}[H]
\begin{center}
   \includegraphics[width=1\columnwidth]{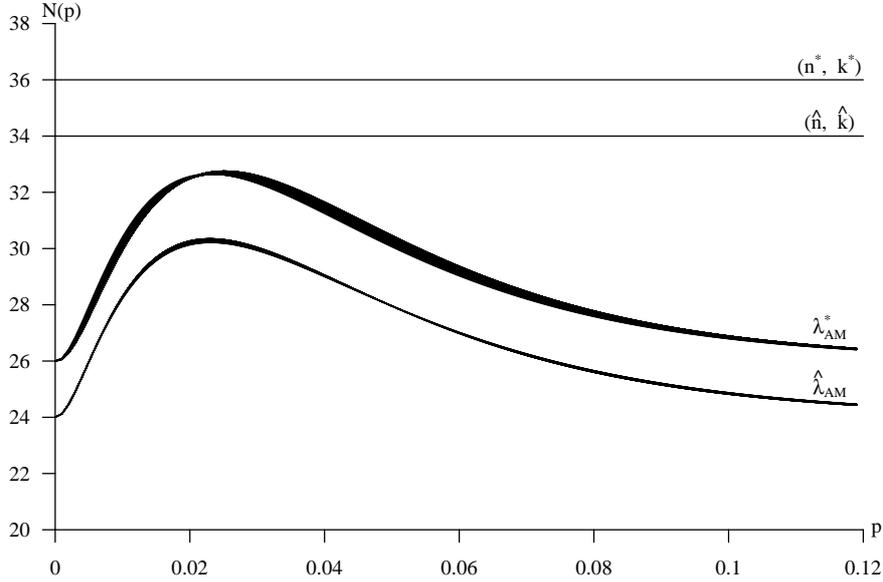}
   \vspace{-10mm}
    \caption{ASN bands for $\lambda^*_{AM}$ and $\widehat{\lambda}_{AM}$ defined by $p_1=0.01$, $p_2=0.06$ and $\alpha=\beta=0.1$}
    \label{ASN3}
\end{center}
\end{figure}
\begin{center}
{\sc References}
\end{center}
{\sc Bruhn-Suhr, M.}, {\sc Krumbholz, W.} (1990). A new variables sampling plan for normally distributed lots with unknown standard deviation and double specification limits. Statistical Papers 31, 195-207.\\\\
{\sc Bruhn-Suhr, M.}, {\sc Krumbholz, W.} (1991). Exact two-sided Lieberman-Resnikoff sampling plans. Statistical Papers 32, 233-241.\\\\
{\sc Hilbert, M.} (2005). Zweifache ASN-Minimax-Variablenprüfpläne
für normalverteiltes Merkmal bei unbekannten Parametern. Thesis, Helmut-Schmidt-Universität, Hamburg.\\\\
{\sc Kolmogorov, A. N.} (1953). Unbiased Estimates. American Mathematical Society Translations 98.\\\\
{\sc Lieberman, G. J.}, {\sc Resnikoff, G. J.} (1955). Sampling Plans for Inspection by variables. Journal of the American Statistical Association 50, 457-516.\\\\
{\sc Vangjeli, E.} (2009). ASN-optimale zweistufige Versionen des Gauß- und t-Tests. Thesis, Helmut-Schmidt-Universität, Hamburg.\\\\
\end{document}